\documentclass[aps,prb,twocolumn,showpacs,superscriptaddress]{revtex4}
\usepackage{graphicx}
\usepackage{bm}
\newcommand{\beq}{\begin{equation}}
\newcommand{\eeq}{\end{equation}}

\begin{document}

\title{
Spontaneous Breaking of Four-Fold Rotational Symmetry in Two-Dimensional Electron Systems as a Topological Phase Transition}
\author{M.~V.~Zverev}
\affiliation{Russian Research Centre Kurchatov
Institute, Moscow, 123182, Russia} 
\affiliation{Moscow Institute of Physics and Technology, Moscow, 123098, Russia}
\author{J.~W.~Clark}
\affiliation{McDonnell Center for the Space Sciences \&
Department of Physics, Washington University,
St.~Louis, MO 63130, USA}
\author{Z.~Nussinov}
\affiliation{McDonnell Center for the Space Sciences \&
Department of Physics, Washington University,
St.~Louis, MO 63130, USA}
\author{V.~A.~Khodel}
\affiliation{Russian Research Centre Kurchatov
Institute, Moscow, 123182, Russia}
\affiliation{McDonnell Center for the Space Sciences \&
Department of Physics, Washington University,
St.~Louis, MO 63130, USA}

\date{\today}

\begin{abstract}
Motivated by recent observations of $C_4$ symmetry breaking in strongly
correlated two-dimensional electron systems on a square lattice, we
analyze this phenomenon within an extended Fermi liquid approach.
It is found that the symmetry violation is triggered by a continuous
topological phase transition associated with exchange of
antiferromagnetic fluctuations.  In contrast to predictions of
mean-field theory, the structure of a part of the single-particle
spectrum violating $C_4$ symmetry is found to be highly anisotropic,
with a peak located in the vicinity of saddle points.
\end{abstract}

\pacs{
71.10.Hf, 
71.27.+a,  
71.10.Ay  
}
\maketitle

\noindent
\section{Introduction}

Experimental studies of strongly correlated two-dimensional (2D)
electron systems have revealed violations of the fundamental
symmetries of time reversal and $C_4$ rotational invariance
inherent in the ground states of these systems on a tetragonal
lattice.\cite{campuzano,ando,kapitulnik, hinkov,mook,taillefer}
Considerable theoretical effort has been aimed at understanding the
nature of these phenomena and identifying their underlying
mechanisms.  Kivelson, Fradkin, and Emery\cite{nematic} were the
first to discuss the case of nematic phase transitions, well before
relevant experimental data became available.  Somewhat later, Yamase
and Kohno \cite{yamase1} (within the $t-J$ model) and Halboth and
Metzner \cite{metzner1} (within the Hubbard model) attributed the
breaking of four-fold symmetry to violation of a Pomeranchuk stability
condition \cite{pom} associated with antiferromagnetic fluctuations.
An analogous result was obtained by Valenzuela and Vozmediano within
an extended Hubbard model.\cite{vozmediano}

As a rule, calculations on the ordered side of the implicated
second-order phase transition are carried out within the mean-field
(MF) approach.\cite{oganesyan,kampf,metzner2,kee,oganesyan2,Fradkin_review}
An effective Hamiltonian containing a separable interaction
$d_2({\bf p}) d_2({\bf p}_1)$ with order parameter
$d_2(p_x,p_y)=\cos p_x - \cos p_y $ is adopted to analyze the
onset of $C_4$ symmetry violation and properties of phases
arising beyond the critical point.  (Momentum components
$p_x,p_y$ are measured in units of the inverse lattice constant.)
This approach has the advantages of transparency and analytical
accessibility. However, it has noteworthy shortcomings, including
sacrifice of translational invariance of the interaction.  Furthermore,
the structure of the relevant order parameter is always postulated
in the MF theory; as a rule the simplest assumption is made
consistent with the type of symmetry breaking being considered.
However, the structure of the new ground state often turns out
to be quite intricate, such that it cannot be properly described in
terms of any single order parameter.  As will be seen, it is just
this situation that emerges in dealing with the $C_4$ symmetry
violation in question.

In the scenario proposed here, the system is considered to be on the
{\it disordered} side of an antiferromagnetic phase transition; hence
the corresponding Pomeranchuk stability condition is not violated.
With the system situated far from the transition point, the fluctuation
exchange is readily analyzed and is too weak to gap out the
single-particle spectrum.  It will be shown, however, that even if
the antiferromagnetic fluctuations are weak, their {\it momentum
dependence} is able to promote a {\it topological} phase transition
associated with disruption of $C_4$ rotational invariance.

In Sec.~II we adopt the Landau-Migdal quasiparticle approach to
investigate $C_4$ symmetry breaking in a 2D electronic system
on a square lattice. A simple model with an infinite-range interaction
function is employed in Sec.~III to analyze a quasiparticle
rearrangement due to antiferromagnetic fluctuations. In Sec.~IV we
present and discuss results of numerical calculations for a more
realistic model having a finite-range interaction.  Sec.~V is
devoted to explanation, within the infinite-range model, of the
arc structure of the Fermi line observed in many high-$T_c$ materials.
Our findings are summarized in Sec.~VI.

\noindent
\section{$C_4$ symmetry breaking within the Fermi liquid approach}

Adopting the Landau-Migdal quasiparticle picture, in which
the physical many-fermion system is viewed as a system of
interacting quasiparticles, the genesis of $C_4$ symmetry
breaking can be investigated based on the fundamental
relation\cite{lan,lanl}
\beq
{\partial\epsilon({\bf p})\over \partial {\bf p}}=
{\partial\epsilon^0_{\bf p}\over \partial {\bf  p}} +{1\over 2}{\rm Tr}
\int {\cal F}_{\alpha\beta,\alpha\beta}({\bf p},{\bf p}_1)
{\partial n({\bf p}_1)\over \partial {\bf p}_1} d\upsilon_1,
\label{lansp}
\eeq
where
$d\upsilon=dp_xdp_y/(2\pi)^2$ is an element of 2D momentum space.
This relation connects the quasiparticle spectrum $\epsilon({\bf p})$
with the quasiparticle momentum distribution
$n({\bf p})=\left[1+\exp\left({(\epsilon({\bf p})-\mu) /T}\right)\right]^{-1}$
through a phenomenological interaction function ${\cal F}$.  This function,
which is defined by  a specific static limit of the quasiparticle
scattering amplitude with initial and final energies on the Fermi
surface,\cite{lan,lanl} depends only on the {\it momenta} ${\bf p}$,
${\bf p}_1$ of the colliding quasiparticles.  Of the two particle-hole
channels relevant to the scattering amplitude ${\cal F}$, denoted
$t$ and $u$ in the Mandelstam's terminology, the transverse $t$
channel carries vital information in the momentum transfer
${\bf q}={\bf p}-{\bf p}_1$, whereas the longitudinal $u$ channel
is silent because the corresponding momentum transfer is close to zero.

In homogeneous matter where total momentum is conserved, the first term
on the right side of Eq.~(\ref{lansp}) is just the bare velocity
${\bf p}/M$, with $M$ the free particle mass.\cite{pit}  In the
presence of a crystal-lattice field, the bare group velocity is
multiplied by a quasiparticle effective charge $e_q({\bf p})$.
However, this modification will be ignored, since it reduces merely
to a renormalization of phenomenological coefficients $t_i$
specifying 2D tight-binding electron spectra
\beq
\epsilon^0_{\bf p} = -2\,t_0\,(\cos p_x
+ \cos p_y)+4\,t_1\,\cos p_x\cos p_y+ \cdots  .
\label{tight}
\eeq

We are concerned specifically with the impact of antiferromagnetic
fluctuations on the electron spectra $\epsilon({\bf p})$ calculated
using Eq.~(\ref{lansp}).  Treatment of the effect of these fluctuations
on the interaction ${\cal F}$ does not encounter difficulties
far from the attendant antiferromagnetic phase transition.
The corresponding fluctuation exchange is adequately addressed
within the Ornstein-Zernike (OZ) approximation, which neglects
scattering of fluctuations.  The part of ${\cal F}$ responsible
for the exchange is then
\beq
{\cal F}^e_{\alpha\beta\gamma\delta}({\bf p},{\bf p}_1)=
\lambda^2 {\bm \sigma}_{\alpha\beta}{\bm \sigma}_{\gamma\delta}
\left[({\bf p}-{\bf p}_1-{\bf Q})^2+\xi^{-2}\right]^{-1}.
\label{sfl}
\eeq
The constant $\lambda$ represents the spin-fluctuation vertex and
${\bf Q}=(\pi,\pi)$ is the antiferromagnetic wave vector, while
$\xi$ is the correlation radius.

Inserting Eq.~(\ref{sfl}) into Eq.~(\ref{lansp}) and evaluating
the spin-fluctuation contribution aided by the identity
$2{\bm \sigma}_{\alpha\beta}{\bm \sigma}_{\gamma\delta}=
3\delta_{\alpha\delta}\delta_{\gamma\beta}-
{\bm \sigma}_{\alpha\delta}{\bm \sigma}_{\gamma\beta}$,
one arrives at
\beq
\epsilon({\bf p})=\epsilon^0_{\bf p} +{3\lambda^2\over 2}
\int { n({\bf p}_1)\over ({\bf p}-{\bf p}_1-{\bf Q})^2+\xi^{-2}}
d\upsilon_1.  \label{spec}
\eeq
The normalization condition 2$\int n({\bf p})d\upsilon=\rho$ determines
the chemical potential $\mu$ consistent with density $\rho$. This approach
to the problem is self-consistent provided the dimensionless parameter
$fN(0)$ is rather small, where $f=(3\lambda^2/ 4\pi)\ln(1/\xi)$ is
a coupling constant and $N(0)\simeq 1/2\pi t_0$ is the density of
states of a 2D electron gas on a square lattice having the
tight-binding spectrum (\ref{tight}).

Direct numerical solution of this 2D nonlinear integral equation
is extremely time-consuming.  If only the component of the
interaction (\ref{sfl}) proportional to $d_2({\bf p})d_2({\bf p}_1)$
is retained, then beyond the point where the corresponding
Pomeranchuk stability condition is violated, one obtains the
ordinary mean-field theory equations. However, this approximation
is quite poor for the interaction (\ref{sfl}), which peaks at
momentum transfer ${\bf q}={\bf Q}$.  Accordingly, the customary
MF scenario must be regarded as vulnerable.

Our approach to the problem stems from this observation:
collapse of collective degrees of freedom associated with
violation of {\it sufficient} conditions\cite{pom} for the
stability of the standard Landau Fermi Liquid (FL) state
is not the only possible scenario for the breakdown of $C_4$
symmetry.  A viable alternative is provided by violation of
a {\it necessary} stability condition.\cite{jetplett2001}
This condition requires that an arbitrary admissible variation
$\delta n({\bf p})$ from the FL quasiparticle momentum
distribution $n_F({\bf p})$, while conserving particle number,
must produce a positive change of the ground-state energy $E_0$,
\beq
\delta E_0=\int\left(\epsilon({\bf p};n_F({\bf  p}))-\mu\right)
\delta n({\bf p})d\upsilon> 0,
\label{necs}
\eeq
where $\epsilon({\bf p};n_F)$ is the spectrum of single-particle
excitations and $\mu$ the chemical potential.

Violation of the condition (\ref{necs}) is accompanied by a change
of the number of roots of the equation
\beq
\epsilon({\bf p},n_F)=\mu,
\label{topeq}
\eeq
which implies a change of the topology of the Fermi surface.
For a thorough development of the concept, see the review by
Volovik.\cite{volrev}  Throughout, we adhere to his rigorous
quantitative definition of topological phase transitions,
as distinguished from looser notions such as transitions
between large and small Fermi surfaces that are also prevalent
in the literature. It should be emphasized that in contrast
to the original Lifshitz description,\cite{lifshitz} the topological
transition under consideration is triggered by the {\it interaction
between quasiparticles} (see also
Refs.~\onlinecite{ks,vol1,noz,zb,shagp,schofield,prb2008,jetplett2009}).

\noindent
\section{Quasiparticle rearrangement within a simplified model}

To gain insight into the essence of this scenario, we
restrict the analysis to zero temperature and simplify the
interaction.  Replacement of the interaction term (\ref{sfl})
by an infinite-range form $\sim\delta({\bf q}-{\bf Q})$
leads directly to the explicit version\cite{jetplett2001}
\beq
\epsilon({\bf p})=\epsilon^0_{\bf p}+fn(\epsilon({\bf p}+{\bf Q}))
\label{noza}
\eeq
of relation (\ref{lansp}), where $f$ is the coupling constant
identified above.  This treatment is analogous to that
adopted by Nozi\`eres\cite{noz} in a study of non-FL behavior
of strongly correlated Fermi systems in the case where forward
scattering in the $t$ channel prevails.  Eq.~(\ref{noza}) can be
derived within a standard variational procedure based on the
formula\cite{jetplett2001}
\beq
E=\int [\epsilon^0_{\bf p}n({\bf p})
+{1\over 2}fn({\bf p})n({\bf p}+{\bf Q})] 2d\upsilon
\label{enan}
\eeq
for the energy $E$ of the model quasiparticle system.  This form
for the energy functional admits a greatly simplified analysis
of the problem due to the partial separation of different
${\bf p}$ channels.

To proceed, we observe first of all that at $T=0$, the posed
rearrangement of the initial standard Landau state can occur only
in those 2D systems where there exist hot spots\cite{pines}---points
situated on the Fermi line and connected by the vector ${\bf Q}$.
Indeed, in systems with small quasiparticle filling, the product
$n({\bf p})n({\bf p}+{\bf Q})$ vanishes for any momentum ${\bf p}$,
so that the ground-state energy is independent of the coupling
constant $f$. The same is true in the case of small hole filling.

In systems with hot spots, the rearrangement occurs due to breaking
of the quasiparticle pairs occupying single-particle states with
momenta ${\bf p}$ and ${\bf p}+{\bf Q}$.  The corresponding
domain ${\cal R}$ (the ``reservoir'') consists of four quasi-rectangles,
each adjacent to one of the saddle points
$(0,\pi),(\pi,0),(0,-\pi),(-\pi,0)$ of the tight-binding
spectrum $\epsilon^0_{\bf p}$. Each of the four elements of
${\cal R}$ is confined between (i) the border of the Brillouin zone,
(ii) the {\it counterpart} of the initial Fermi line, defined by the
equation $\epsilon^0_{{\bf p}+{\bf Q}}=\mu$, and (iii) two segments
of the Fermi line embracing the given saddle point.

\begin{figure}[t]
\includegraphics[width=0.98\linewidth,height=1.\linewidth]{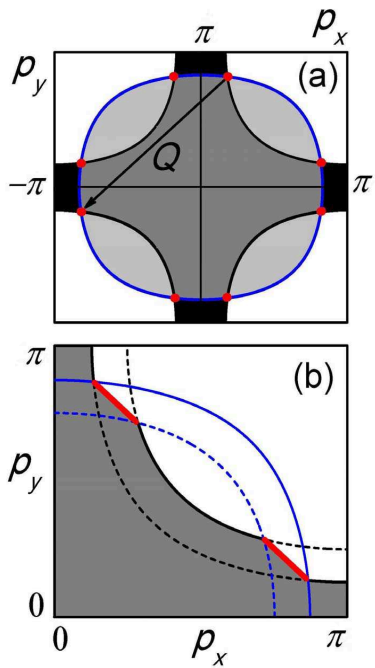}
\caption{Color online.
Panel (a): Fermi line (black) and its counterpart (blue(gray)) for the bare
tight-binding spectrum $\epsilon^0_{\bf p} = -2\,t_0\,(\cos p_x
+ \cos p_y)+4\,t_1\,\cos p_x\cos p_y$, with $t_1/t_0=0.45$. The
reservoirs ${\cal R}$ are colored in black, and the lenses ${\cal L}$, in
light gray.  The hot spots connected with each other by the vector
${\bf Q}$ are symbolized by red (gray) dots. Panel (b): Fermi line for the
model assuming the infinite-range interaction function
$f({\bf q})=(2\pi)^2 f\delta({\bf q}-{\bf Q})$, with  $fN(0)=0.13$.
Hot lines are drawn in red (gray).  Fermi lines for the bare
tight-binding spectrum $\epsilon^0_{\bf p}$ and its counterpart are
shown as green (light gray) and blue (gray) lines, respectively.
}
\label{fig:zero}
\end{figure}

Quasiparticles move out the domain ${\cal R}$ to resettle in
a region ${\cal L}$ where all pairs of
single-particle states connected by the vector ${\bf Q}$ are empty.
The region ${\cal L}$ comprises four ``lenses,'' situated between
neighboring hot spots and bounded by the initial Fermi line
and its counterpart (see panel (a) of Fig.~\ref{fig:zero}).
The transfer of one quasiparticle from ${\cal R}$ to ${\cal L}$
produces a gain in energy which is just the coupling
constant $f$ minus the loss $\tau$ of kinetic energy.
Its minimum $\tau_{\rm min}$ is attained when a quasiparticle,
vacating a state in ${\cal R}$ with momentum ${\bf p}$, occupies
in ${\cal L}$ a state of lowest energy, given by the chemical
potential, so that $\tau_{\rm min}= \mu - \epsilon_{\bf p}^0$.
Therefore the rearrangement is favorable provided
$\epsilon^0_{\bf p}-\mu+ f\geq 0$.

An alternative process involves transfer of the quasiparticle
counterpart, which has momentum ${\bf p}+{\bf Q}$. In this case, the
rearrangement occurs provided $\epsilon^0_{{\bf p}+{\bf Q}}-\mu+f\geq 0$.
The choice between the two options is decided by comparing the
corresponding energies.  The boundary at which one behavior gives
way to the other is determined by the relation
$\epsilon^0_{\bf p}=\epsilon^0_{{\bf p}+{\bf Q}}$.  Since the
straight line so defined is part of the {\it new} Fermi line, we
infer that the rearrangement has converted the original, isolated
hot spot into a {\it continuous straight line} of hot spots,
i.e., a hot line (HL) (see panel (b) Fig.~\ref{fig:zero}).

These results imply that quasiparticles are swept from a certain
subdomain ${\cal S}$ of ${\cal R}$ consisting of eight approximately
trapezoidal strips.  The boundaries of a given strip are traced
on three sides by (respectively) the initial Fermi line, the border
of the Brillouin zone, and a line geometrically similar to the
initial Fermi line but shifted into the domain ${\cal R}$ (see
Fig.~\ref{fig:zero}).  The strip's fourth side (red on-line) is just
the hot line.  This solution is self-consistent: any single-particle
state with momentum ${\bf p}\in {\cal S}$ has its counterpart, with
momentum ${\bf p}+{\bf Q}$, located outside ${\cal S}$, and this state
is occupied, so that Eq.~(\ref{noza}) is fulfilled.  Transparently,
the new momentum distribution {\it does not} violate $C_4$ symmetry.

Defining the strip energy width $W_s$ of the region ${\cal S}$ as
the maximum of the initial hole energy $|\epsilon^0_{\bf p}-\mu_i|$
consistent with the rearrangement, one has $W_s=D_i-D_f$, where $2D_i$
(respectively, $2D_f$) is the minimum energy distance between the
segments of the initial (final) Fermi line situated in different
half-planes.  On the other hand, one finds $W_s=f-(\mu-\mu_i)$,
where $\mu_i$ is the initial chemical potential.  To estimate the
strip width $W_s$ and the difference $\mu-\mu_i$, both proportional
to $f$, we (i) approximate the Fermi velocity
${{\bf v}^0({\bf p})}=\left(\partial \epsilon^0_{\bf p}/
\partial {\bf p}\right)_0$ on the Fermi line in terms of two parameters,
namely its average magnitudes $v_l^0$ and $v_s^0$ in the lens and strip
regions, respectively, and (ii) invoke the coincidence of the chemical
potential with the Fermi energy that is intrinsic to Landau theory.
In the lens region ${\cal L}$ one then has $\mu-\mu_i=v_l^0w_l$, where
$w_l$ is the momentum width of the lens filling.  In the domain ${\cal S}$,
one obtains the analogous formula $W_s \equiv f-(\mu-\mu_i)=w_sv_s^0$.
Particle-number conservation implies that $w_l l_l = 2 w_s (l_s-w_s/2) $,
where $l_s$ is the strip length, $l_l$ is the lens length, and
$w_s=W_s/v^0_s$ is the momentum width of the strip.
Upon elimination of $w_s$ and $w_l$ from these relations,
we arrive finally at
\beq
\mu-\mu_i= {2fv_l^0l_s\over 2v_l^0l_s+v_s^0l_l},
\label{mu}
\eeq
for small $w_s$.

As long as all the saddle points remain occupied, $C_4$ symmetry is
preserved.  However, as the electron density $\rho$ decreases, the
distance between the new Fermi line and the saddle points shrinks.  At
a critical density $\rho_c$, or equivalently, at the critical constant
$f_c$ where two segments of the Fermi line that cross the same
boundary of the Brillouin zone {\it merge} at the saddle point,
the number of solutions of Eq.~(\ref{topeq}) certainly drops,
thereby signaling a {\it topological phase transition}. In the
critical situation one has $D_f=0$, or equivalently $W_s=D_i$.
The trapezoidal shape ${\cal S}$ then becomes triangular, and we have
\beq
D_i={f_cl_l\over v_l^0l_s+v_s^0l_l}\simeq f_c.
\label{mut}
\eeq
Using this result, the critical value $F_c$ of the dimensionless
constant $F=fN(0)$ is given by $F_c={D/2\pi t_0}$.  Assuming the ratio
$D/t_0$ to be small, we thus have $F_c \ll 1$, which implies that the
derivative $\partial\Sigma({\bf p},\varepsilon)/\partial \varepsilon$
remains small, i.e., that the $\varepsilon$-dependence of the
mass operator $\Sigma({\bf p},\varepsilon)$ is moderate.\cite{qcp}
Under these conditions, the generation of new branches of the
single-particle spectrum $\epsilon({\bf p})$, such as the small pockets
of the Fermi surface suggested to explain magnetic oscillations
in the pseudogap regime,\cite{smpoc} is questionable.

Beyond the transition point (e.g.\ at $\rho<\rho_c$), $C_4$ symmetry
is necessarily broken.  Suppose, conversely, that it is preserved.
Then all the saddle points must then be emptied simultaneously,
implying that every rearranged saddle point energy $\epsilon_s(\rho)$
{\it exceeds} the chemical potential $\mu(\rho)$.  But according to
Eq.~(\ref{noza}), the interaction contribution to $\epsilon_s$ vanishes
when all the saddle points are emptied.  Hence the saddle-point
energy $\epsilon_s(\rho)$ must coincide with the corresponding bare
value $\epsilon^0_s(\rho)$, implying that $\epsilon^0_s(\rho)>\mu(\rho)$.
However, if the difference $\rho_c-\rho$
is small, then without fail $\epsilon^0_s(\rho)<\mu_i(\rho)$. Thus
a contradiction is encountered, since it follows from Eq.~(\ref{mu})
that $\mu_i(\rho) < \mu(\rho)$.  This deadlock is resolved if,
beyond the critical point, only {\it one} of two neighboring
saddle points is emptied, with the second remaining occupied.
Such a solution is indeed consistent with Eq.~(\ref{noza}).

\noindent
\section{Numerical results with realistic interaction}

\begin{figure}[t]
\includegraphics[width=0.89\linewidth,height=1.\linewidth]{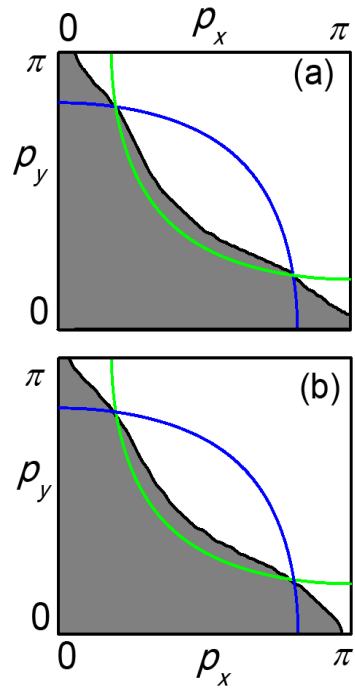}
\caption{Color online.
Fermi lines for the model assuming the finite-range
interaction function $f({\bf q}){=}f_a/(({\bf q}{-}{\bf Q})^2{+}\xi^{-2})$,
with $\xi{=}30$. Panel (a): $f_aN(0){=}0.32$; $C_4$ symmetry is not broken.
Panel (b): $f_aN(0){=}0.48$; one of the two solutions with spontaneously
broken $C_4$ symmetry is shown.  Only the first quadrant of the
Brillouin zone is drawn since neither $p_x{\to}-p_x$ nor
$p_y{\to}-p_y$ reflection symmetry is broken. Fermi lines for the
bare tight-binding spectrum $\epsilon^0_{\bf p}$ and its counterpart
are shown as green (light gray) and blue (gray) lines respectively.
}
\label{fig:2_3}
\end{figure}

\begin{figure}[t]
\includegraphics[width=0.9\linewidth,height=0.75\linewidth]{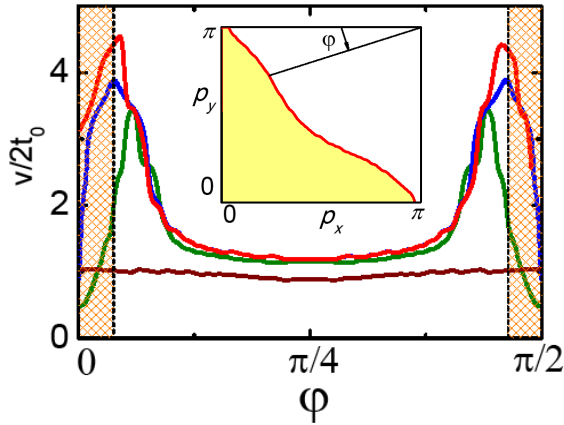}
\caption{Color online.
Fermi-velocity magnitudes $v_F=|\partial\epsilon({\bf p})/\partial
{\bf p}|$ (in units of $2t_0$), evaluated along the Fermi line as
a function of the angle $\varphi$ defined in the inset, for different
single-particle spectra $\epsilon({\bf p})$. Results are shown
for the bare tight-binding model with the same parameter
choice as in Fig.~\ref{fig:zero} (brown (dark gray) line) and for the
Fermi-liquid-theory model of Fig.~\ref{fig:2_3} at $f_aN(0)=0.32$,
$T=10^{-4}$ (green (light gray) line); $f_aN(0)=0.48$, $T=10^{-4}$
(red (black) line); and $f_aN(0)=0.48$, $T=10^{-2}$ (blue (gray) line).
Broken $C_4$-symmetry of the solid/red curve with respect to $x-y$
exchange is manifested by its different behavior in the two
shaded areas close to the saddle points.
}
\label{fig:xyvelo_c}
\end{figure}

The results we have derived for the simple model based on an
infinite range interaction $\sim \delta({\bf q} - {\bf Q})$
are in agreement with those obtained from numerical calculations
performed for the more realistic interaction (\ref{sfl}) and
displayed in Figs.~\ref{fig:2_3} and (\ref{fig:xyvelo_c}).  Some
complications associated with the finite correlation radius of the
interaction (\ref{sfl})  will be considered below, but first we
examine the results of the extended Fermi-liquid theory in comparison
with corresponding predictions of MF theory.  The MF single-particle
spectrum coincides with a bare spectrum before the transition point
is reached, while beyond the transition it receives a correction
$\delta\epsilon_{\rm MF}({\bf p})=\eta (\cos p_x-\cos p_y)$, with
the order parameter $\eta$ taking the same value throughout the Brillouin
zone.  The Fermi line calculated within the extended FL approach deviates
substantially from that predicted by MF theory.  In particular,
upon comparing the upper and lower panels of Fig.~\ref{fig:2_3},
we see that in the lens domain the location of the Fermi line
remains almost unchanged as the system passes through the transition
point.  Indeed, this behavior also prevails over a significant portion
of the HL region away from the saddle points.  In other words,
beyond the point where $C_4$ symmetry is lost, the associated
rearrangement of the Fermi surface occurs only in the immediate
vicinity of the saddle points---in a sharp contrast to what
is found in MF theory.

Analogous conclusions follow from a study of Fig.~\ref{fig:xyvelo_c},
where the Fermi velocity calculated on the basis of Eq.~(\ref{lansp})
is plotted.  It is seen that the correction to the the bare Fermi
velocity $v^0_F$ stemming from antiferromagnetic correlations as
described by (\ref{sfl}) remains smooth and small except in the
HL region, where it soars upward.

Such behavior of the Fermi velocity $v_F$, which persists through
the transition point, can be elucidated by analyzing the Landau
relation (\ref{lansp}).  First, we observe that the overwhelming
contributions to the integral in this relation come from the HL
region; otherwise there is no appreciable overlap between the
peak in the interaction function and the $\delta$ peak in the
derivative $\partial n({\bf p})/\partial {\bf p}$.   To proceed
further, we introduce a new set of orthogonal momentum coordinates
$p_t,p_n$, with the axis $p_t$ directed along the HL and the axis
$p_n$ perpendicular to it.  In the HL region we then have
$dp_xdp_y=dp_ndp_t$ and $dn({\bf p})/dp_n=\pm\delta(p_n)$, the
sign of the derivative being positive in the left half-plane and
negative otherwise.  This alternate sign is responsible for the
vanishing of the group velocity at the saddle points, through
interference of the contributions to the integral term in
Eq.~(\ref{lansp}) from neighboring segments of the Fermi line
situated in the two half-planes.  The distance between these
segments (as defined in Sec.~III) is $2D_f$.  If the inverse
correlation radius $\xi^{-1}$, which measures the radius of the
spin-interaction term (\ref{sfl}) in momentum space, turns out to
be so small that $\xi^{-1}\leq D_f$, then the two contributions
cease to interfere, and the elevation of the HL value of the Fermi
velocity is readily estimated as
\beq
v_F(p_t)\simeq (2\pi)^{-2}\xi^{-1}.
\label{vf}
\eeq
These conclusions are in agreement with the results for the model
with finite-range interaction presented in Fig.~\ref{fig:xyvelo_c}.
At the same time, the estimate (\ref{vf}) is in agreement with the
jump of the single-particle spectrum $\epsilon({\bf p})$ on crossing
the HL, found for the simple model with $\delta({\bf q}-{\bf Q})$
interaction and implying an infinite value of the model's HL group
velocity.  The above considerations demonstrate that the FL
rearrangement of the ground state leading to the phenomenon of
$C_4$ symmetry violation has little in common with the
rearrangement  predicted by conventional MF theory based on
the single order parameter $d_2$.

The analysis can be made more informative by focusing on
the difference $D(p_x,p_y)=\epsilon(p_x,p_y)-\epsilon(p_y, p_x)$
and its integral $D$ over the intermediate momenta $p_x,p_y$.
Both quantities vanish on the disordered side of the phase
transition, and beyond the transition point it is straightforward
to evaluate $D$ by means of Eq.~(\ref{spec}).  For $D\to 0$, one
can make use of the formula
$n(p_x,p_y)-n(p_y, p_x)=(dn({\bf p}/d\epsilon({\bf p})D(p_x,p_y)$
to recast this equation in a form
\beq
D({\bf p})=-\int {\cal F}({\bf p},{\bf p}_1){\partial n({\bf p}_1)
\over \partial \epsilon({\bf p}_1)} D({\bf p}_1){d^2p_1\over (2\pi)^2}
\label{ord}
\eeq
equivalent to the Pomeranchuk stability condition, whose violation
is a prerequisite for the MF description of $C_4$ symmetry breaking.
From the preceding discussion, we infer that if a nontrivial solution
of Eq.~(\ref{ord}) exists, it must be anisotropic, with a peak
located in the HL domain and having a width of order $\xi^{-1}$.
Such a structure of the order-parameter function $D({\bf p})$ is
quite unlike that adopted in conventional MF theory of the
observed $x-y$ symmetry violation.
In evaluating the integral in Eq.~(\ref{ord}) we employ the relation
$\partial n({\bf p})/\partial \epsilon({\bf p})=(dn(p_n)/dp_n)/v_F$.
Referring to the above derivation of the estimate (\ref{vf}), it is seen
that the relevant value of the group velocity is $v_F \simeq \xi$,
as long as $\xi \geq (D_i-W_s)^{-1}$.  The $\xi$ dependence of the
integral is then effectively nullified, precluding nontrivial
solutions of Eq.~(\ref{ord}).

Nontrivial solutions of Eq.~(\ref{ord}) can in fact emerge before the two
neighboring pieces of the Fermi line meet each other at the saddle
point, provided $\xi\leq  (D_i-W_s)^{-1}$.  In this case, the characteristic
value of the Fermi velocity drops somewhat, thereby enhancing the integral.
Whether this enhancement is sufficient for the violation of the
Pomeranchuk stability condition will be decided in a more intensive
round of numerical calculations.

It is worth emphasizing that the situation underlying the violation
of $C_4$ symmetry in systems in which the Fermi surface comes close
to van Hove points is not specific to either the MF treatment or
our analysis.  In fact, the effective Stoner factor, which determines
the enhancement of the effective field acting on a particle in matter,
is proportional to the product of the interaction strength and the
density of states. The latter diverges at a van Hove point, and
hence the corresponding Stoner factor diverges as well, {\it
independently of the shape of the order parameter}. The crucial
point of distinction is as follows. In MF theory, which reasonably
exploits the enhancement of the density of states near the
van Hove points and an order parameter $d_2({\bf p})$ having
the needed symmetry, the effective field stretches over the
whole Brillouin zone in accordance with the chosen shape of the
order parameter. In our approach based on exchange of
antiferromagnetic fluctuations between electrons, it is instead
the shape of the exchange interaction that governs the behavior
of the effective field.
Since this field dies out at rather small distances from the saddle points, the topological rearrangement of the Fermi surface violating $C_4$ symmetry occurs only in the regions close to these points.

Let us now identify inherent properties of the interaction
function ${\cal F}$ responsible for violation of $C_4$ symmetry
and more generally for topological transitions.  In homogeneous
matter, it is well understood that topological phase transitions
are characterized by a change in the number of sheets of the
Fermi surface.\cite{prb2008}  In 2D electron systems on a square
lattice, topological transitions are of much the same character.
The salient common feature here is that {\it no symmetry is violated},
provided that a local rearrangement of the quasiparticle
momentum distribution leads to dominance of {\it forward} scattering
in the $t$ channel referred to the momentum transfer ${\bf q}$
specifying ${\cal F}({\bf q})$.  On the other hand, in the case of
antiferromagnetic fluctuations {\it backward} scattering prevails.
Then, at the transition point, quasiparticles leaving the vicinity
of one saddle point may move into the vicinity of a neighboring
saddle point.  Thus the sheet number remains unchanged; instead,
the symmetry of the ground state is violated.

\noindent
\section{Arc structure of the Fermi line}

The model we have developed may also have a bearing on the emergence of
the arc structure of the Fermi line observed in many high-$T_c$
materials.  If we consider pairing based on the interaction (\ref{sfl}),
then Eq.~(\ref{enan}) must be supplemented by a pairing term
$(f/2)\kappa({\bf p})\kappa({\bf p}+{\bf Q})$,\cite{noz} where
$\kappa({\bf p}){=}\langle a^+({\bf p})a^+(-{\bf p})\rangle$
is a superfluid density.  With this modification, Eq.~(\ref{noza})
as written remains unchanged, but the quasiparticle occupation
number $n({\bf p})$ acquires the BCS form
$n({\bf p}){=}1/2{-}\epsilon({\bf p})/2E({\bf p})$, with quasiparticle
energy $E({\bf p}){=}\left[\epsilon^2({\bf p})
{+}\Delta^2({\bf p})\right]^{1/2}$.
The additional equation
\beq
\Delta({\bf p})=-f {\tanh (E({\bf p}+{\bf Q})/2T)\over 2E({\bf p}+{\bf Q})}
\Delta({\bf p}+{\bf Q})
\label{gap0}
\eeq
determines the gap function $\Delta({\bf p})$.
In advance of the topological phase transition, where $C_4$ symmetry
is preserved, a standard nonzero solution of Eq.~(\ref{gap0}) has
the property $\Delta({\bf p})= -\Delta({\bf p}+{\bf Q})$ exhibited
by $D$ pairing, and we find
\beq
{E({\bf p})E({\bf p}+{\bf Q})
\over \tanh (E({\bf p})/2T)\tanh (E({\bf p}+{\bf Q})/2T)}={f^2\over 4}.
\label{gap}
\eeq

As seen from Eq.~(\ref{gap}), the associated gap $E_{\rm min}$ in
the single-particle spectrum is suppressed near the diagonals of
the Brillouin zone, where
\beq
E_{\rm min}(T=0)\sim {f^2\over 4W_l},
\label{eml}
\eeq
$W_l$ being the total energy lens width.
On moving along the Fermi line toward the hot line where one
has $E({\bf p})= E({\bf p}+{\bf Q})$, the gap soars upward,
with Eq.~(\ref{gap}) yielding
\beq
E({\bf p}, T=0) \simeq {f\over 2}.
\label{emh}
\eeq
It is important to note
that in the HL region itself, the gap value is markedly suppressed,
because Eq.~(\ref{noza}) tells us that $|\epsilon({\bf p})|\simeq f$ in
a significant part of this region, which is incompatible with
Eq.~(\ref{emh}).  This indicates
that pairing has little impact on the violation of $C_4$ symmetry,
which primarily involves the immediate vicinities of the hot lines.

\noindent
\section{Conclusion}

In summary, we have addressed the problem of $C_4$-symmetry violation
in electron systems on a square lattice within a self-consistent Fermi
liquid approach, assuming that the Landau interaction describes
the exchange of antiferromagnetic fluctuations, which is treated within
the Ornstein-Zernike  approximation.  We have demonstrated that
as the strength of this interaction builds up, the distance between
saddle points and the Fermi line shrinks, eventually generating
a quantum critical point of a new type, at which a {\it continuous}
topological phase transition triggers the violation of $C_4$ symmetry. The
group velocity becomes finite again once the transition point is
passed.  Thus, the properties of the electron system are governed
by Fermi-liquid theory throughout the vicinity of the proposed
quantum critical point, implying that magnetic oscillations should
be observed on both the sides of the topological transition, in
agreement with recent measurements.\cite{mackenzie}

\noindent
\acknowledgments

We express our gratitude to V.~Yakovenko and H.~Yamase
for comprehensive discussion of key points and also
thank A.~S.~Alexandrov, A.~Balatsky, E.~Fradkin, A.~Mackenzie,
and V.~Shaginyan for fruitful discussions.  This research was
supported by the McDonnell Center for the Space Sciences, by
Grants Nos.~2.1.1/4540 and NS-7235-2010.2 from the Russian Ministry
of Education and Science, and by Grant No.~09-02-01284 from the
Russian Foundation for Basic Research.


\smallskip

\begin{thebibliography}{99}

\bibitem{campuzano} A.\ Kaminski, S.\ Rosenkranz, H.\ W.\ Fretwell,
J.\ C.\ Campuzano, Z.\ Li, H.\ Raffy, W.\ G.\ Cullen, H.\ You,
C.\ M.\ Varma, and H.\ H.\ Hoehst, Nature {\bf 416}, 610 (2002).

\bibitem{ando} Y.\ Ando, K.\ Segawa, S.\ Komiya, and A.\ N.\ Lavrov, Phys.\ Rev.\ Lett.\ {\bf 88}, 137005 (2002).

\bibitem{kapitulnik} J.\ Xia, E.\ Schemm, G.\ Deutscher, S.\ A.\ Kivelson, D.\ A.\ Bonn,
W.\ N.\ Hardy, R.\ Liang, W.\ Siemons, G.\ Koster, M.\ M.\ Fejer, and
A.\ Kapitulnik,  Phys.\ Rev.\ Lett.\ {\bf 100}, 127002 (2008).

\bibitem{hinkov} V.\ Hinkov, D.\ Haug, B.\ Fauque, P.\ Bourges, Y.\ Sidis, A.\ Ivanov,
C.\ Bernhard, C.\ T.\ Lin, and B.\ Keimer, Science {\bf 319}, 597 (2008).

\bibitem{mook} H.\ A.\ Mook, Y.\ Sidis,  B.\ Fauque, V.\ Baledent, and P.\ Bourges,
Phys.\ Rev.\ B {\bf 78}, 020506(R) (2008).

\bibitem{taillefer} K.\ Daou, J.\ Chang, D.\ LeBoeuf, O.\ Cyr-Choiniere, F.\ Laliberte,
N.\ Doiron-Leyraud, B.\ J.\ Ramshaw, R.\ Liang, D.\ A. Bonn, W.\ N.\ Hardy, and L.\ Taillefer, Nature {\bf 463}, 519 (2010).

\bibitem{nematic}
S.\ A.\ Kivelson, E.\ Fradkin, and V.\ J.\ Emery, Nature {\bf 393}, 550 (1998).

\bibitem{yamase1}
H.\ Yamase and H.\ Kohno, J.\ Phys.\ Soc.\ Jpn {\bf 69}, 2151 (2000).

\bibitem{metzner1}C.\ J.\ Halboth and W.\ Metzner,
Phys.\ Rev.\ Lett.\ {\bf 85}, 5162 (2000).

\bibitem{pom} I.\ Ya.\ Pomeranchuk, Sov.\ Phys.\ JETP {\bf 8}, 361 (1959).

\bibitem{vozmediano} B.\ Valenzuela and M.\ A.\ H.\ Vozmediano, Phys.\ Rev.\ B {\bf 63}, 153103 (2001).

\bibitem{oganesyan} V.\ Oganesyan, S.\ A.\  Kivelson, and E.\ Fradkin, Phys.\ Rev.\ B  {\bf 64}, 195109 (2001).

\bibitem{kampf} A.\ P.\ Kampf and A.\ A.\ Katanin, Phys.\ Rev.\ B {\bf 67}, 125104 (2003).

\bibitem{oganesyan2} I.\ Khavkine, C.\ H.\ Chung,  V.\ Oganesyan, H.\ Y.\ Kee, Phys. Rev. B {\bf 70}, 155110 (2004).

\bibitem{metzner2} A.\ Neumayr, W.\ Metzner, Phys.\ Rev.\ B {\bf 67},
035112 (2003).

\bibitem{kee} H.\ Y.\ Kee, E.\ H.\ Kim, and C.\ H.\ Chung, Phys.\ Rev.\ B  {\bf 68}, 245109 (2003).

\bibitem{Fradkin_review} E.\ Fradkin, S.\ A.\ Kivelson, M.\  J.\ Lawler, J.\ P.\ Eisenstein, and A.\ P.\ Mackenzie, arXiv:0910.4166

\bibitem{lan} L.\ D.\ Landau,  Sov.\ Phys.\ JETP {\bf 3}, 920 (1957);
{\bf 8}, 70 (1959).

\bibitem{lanl} L.\ D.\ Landau and E.\ M.\ Lifshitz, {\it Course of Theoretical Physics}, Vol.~5, {\it Statistical Physics}, 3rd edition (Nauka, Moscow, 1976; Addison-Wesley, Reading, MA, 1970).

\bibitem{pit} L.\ P.\ Pitaevskii, Sov.\ Phys.\ JETP {\bf 10}, 1267 (1960).

\bibitem{jetplett2001} M.\ V.\ Zverev, V.\ A.\ Khodel, and J.\ W.\ Clark,
JETP Lett. {\bf 74}, 46 (2001).

\bibitem{volrev} G.~E.~Volovik, Springer Lecture Notes in Physics
{\bf 718}, 31 (2007) [cond-mat/0601372].

\bibitem{lifshitz} I.\ M.\ Lifshitz, Sov.\ Phys.\ JETP {\bf 11}, 1130 (1960).

\bibitem{ks} V.~A.~Khodel and V.~R.~Shaginyan, JETP Lett. {\bf 51}, 553
(1990).

\bibitem{vol1}G.~E.~Volovik, JETP Lett.\ {\bf 53}, 222 (1991).

\bibitem{noz} P.~Nozi\`eres, J.~Phys.~I France {\bf 2}, 443 (1992).

\bibitem{zb}  M.~V.~Zverev and M.~Baldo, JETP {\bf 87}, 1129 (1998);
J.\ Phys.: Condens.\ Matter {\bf 11},  2059 (1999).

\bibitem{shagp} S.~A.~Artamonov, V.~R.~Shaginyan, and Yu.~G.~Pogorelov,
JETP Lett.\ {\bf 68}, 942 (1998).

\bibitem{schofield} J.\ Quintanilla and A.\ J.\ Schofield, Phys.\ Rev.\ B
{\bf 74}, 115126 (2006).

\bibitem{prb2008} V.\ A.\ Khodel, J.\ W.\ Clark, and M.\ V.\ Zverev,
Phys.\ Rev.\ B {\bf 78}, 075120 (2008); and references cited therein.

\bibitem{jetplett2009}  V.\ A.\ Khodel, J.\ W.\ Clark, and M.\ V.\ Zverev, JETP Lett. {\bf 87}, 693 (2009); arXiv:0904.1509.

\bibitem{pines} D.\ Pines, Physica C {\bf  282-287},  273 (1997);
A.\ V.\ Chubukov, Europhys.\ Lett.\ {\bf 44}, 655 (1998).

\bibitem{qcp} V.\ A.\ Khodel, J.\ W.\ Clark, and M.\ V.\ Zverev, JETP Lett.\ {\bf 90}, 693 (2009).

\bibitem{smpoc} Y.\ Qi and S.\ Sachdev, Phys.\ Rev.\ B {\bf 81} 115129 (2010);
M.\ Khodas and A.\ M.\ Tsvelik, Phys.\ Rev.\ B {\bf 81} 155102 (2010).

\bibitem{mackenzie} J.-F.\ Mercure, S.\ K.\ Goh, E.\ C.\ T.\ O'Farrell, R.\ S.\ Perry, M.\ L.\ Sutherland, A.\ W.\ Rost, S.\ A.\ Grigera, R.\ A.\ Borzi, P.\ Gegenwart, A.\ P.\ Mackenzie, Phys.\ Rev.\ Lett.\ {\bf 103}, 176401 (2009).
\end{thebibliography}
\end{document}